\documentclass[prl,twocolumn]{revtex4}

\usepackage{graphicx}

\newcommand{\cms}{Co$_2$MnSi}
\newcommand{\bto}{BaTiO$_3$}

\begin{document}

\title{Interface effects at a half-metal/ferroelectric junction}

\author{Kunihiko Yamauchi$^1$, Biplab Sanyal$^2$ and Silvia Picozzi$^1$}

\affiliation{Consiglio Nazionale delle Ricerche - Istituto Nazionale per la Fisica della Materia (CNR-INFM), CASTI Regional Lab, I--67010 Coppito (L'Aquila),
Italy}
\affiliation{Theoretical Magnetism Group, Department of Physics, Uppsala University, Box-530, SE-75121, Sweden}

\begin{abstract}

Magnetoelectric effects are investigated ab-initio at the interface between  half-metallic and ferroelectric prototypes:  Heusler \cms\ and perovskite \bto. For the Co-termination ferroelectricity develops in \bto\ down to nanometer thicknesses, whereas for the MnSi-termination a paraelectric and a ferroelectric state energetically compete, calling for a full experimental control over the junction atomic configuration whenever a ferroelectric barrier is needed. Switch of the electric polarization largely affects magnetism in \cms, with magnetoelectric coupling due to electronic hybridization at the MnSi termination and to structural effects at the Co-termination.  Half-metallicity is lost at the interface, but recovered already in the subsurface layer.
\end{abstract}
\maketitle

The control over the magnetic (ferroelectric) properties via an electric (magnetic) field in
magnetoelectric (ME) materials has enormous implications from the technological point of view  in future nanoscale devices.\cite{nicolanatmat} Since a prototype {\em single-phase} ME multiferroic (MF) material  
suitable for industrial applications has not emerged so far\cite{mostovoy,wang,slv}, the alternative of
 {\em two-phase} or {\em composite} system, where a ferroelectric (FE) or MF compound is interfaced 
with a ferromagnetic material, is becoming increasingly  popular.\cite{scott,schlom,tsymbalprl,gajek}

Here, we present a Co$_{2}$MnSi/BaTiO$_{3}$ layered nanostructure
as a half-metallic/ferroelectric  junction showing  remarkable magneto-electric effects.
The predicted full spin polarization\cite{iosif,heubulk} and the high Curie temperature (905 K), make Co$_{2}$MnSi (CMS) an attractive candidate as electrode for spin-injection in magnetic tunnel junctions.\cite{yamamoto}
In parallel, BaTiO$_{3}$ (BTO) is a prototypical perovskite-like FE oxide\cite{cohen}	
which also shows high Curie temperature ($\sim$ 400 K).  The choice of the materials is also supported by the negligible  lattice constant mismatch ($\sim$0.1$\%$) between CMS (110) and BTO (100) .

\begin{figure}[htbp]
\begin{center}
\vspace{1.cm} \hspace{0.5cm}
\includegraphics[width=70mm]{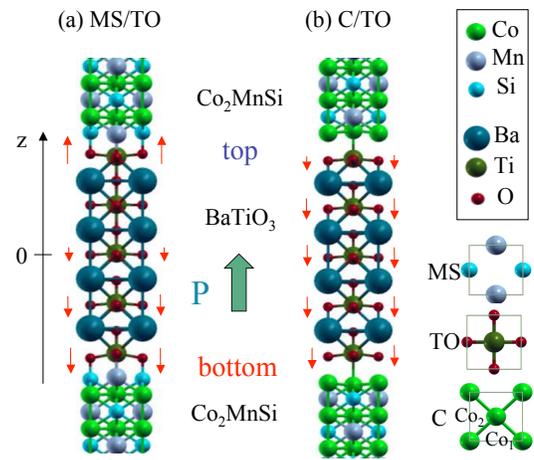}
    \caption{Atomic structure of Co$_{2}$MnSi/BaTiO$_{3}$/Co$_{2}$MnSi junction with (a) MnSi-TiO$_{2}$  and (b) Co$_{2}$-TiO$_{2}$  termination. Red arrows indicate the direction of displacement of O atoms relative to Ti atoms. In both cases, the total electric polarization points up. In the right-side, the top view of the interface layers is also shown.}
      \label{fig1}
\end{center}
\end{figure}
 
First-principles density functional calculations within the generalized gradient approximation (GGA)\cite{pbe}
were performed with the VASP\cite{vasp} code, using
an energy cutoff of 400 eV for the plane wave expansion
and $8\times 8\times 1$ grid for the $k$-point sampling.\cite{mp}
CMS  (110) and BTO (100) layers were stacked periodically in a supercell approach   
to simulate the [001]-ordered Co$_{2}$MnSi/BaTiO$_{3}$ interface (cfr Fig.\ref{fig1}).  In order
to study the interplay between ferroelectric/magnetic properties and the interface structure, 
both Co$_{2}$- and MnSi- terminations in CMS layers were simulated. 
As for the FE side, only TiO$_{2}$- termination was considered, 
as this is believed to be BTO's most stable termination with magnetic materials 
(cfr Fe/BaTiO$_{3}$ \cite{tsymbalprl}). Various thicknesses of the FE barrier were considered, so that 
our supercells can be described as MnSi-(Co$_{2}$MnSi)$_{4}$-TiO$_{2}$-(BaO-TiO$_{2}$)$_{m}$ (denoted as MS/TO terminated) and (Co$_{2}$MnSi)$_{4}$-Co$_{2}$-TiO$_{2}$-(BaO-TiO$_{2}$)$_{m}$ (denoted as C/TO terminated) where $m=2,4,6,8,10$. Our structural properties were obtained as follows:
first, we minimized the total energy by relaxing all the atomic positions 
{\em and} by changing the Bravais lattice $c$-length for the supercells with $m=2$ and $4$. 
Then, we increased the number of (BaO-TiO$_{2}$) layers, 
by rigidly adding to the  $m=4$ supercell, one or more BTO unit cells (with a calculated equilibrium lattice constant $c = 4.08$ \AA)
in the insulating side. Therefore, 
for larger supercells with $m=6, 8, 10$, the internal degrees of freedom were fully relaxed, keeping the $c$-lengh fixed.
This procedure was performed for each atomic configuration
with and without mirror symmetry imposed on the center of BTO layers, in order
to compare the total energy between paraelectric (PE) and FE states.
We point out that, after atomic relaxation, both PE and FE atomic configurations are obtained for each interface, except for $m=2$ case where the system only shows PE state.
This indicates that the critical thickness  for ferroelectricity of BaTiO$_{3}$ layers in this system is less than 1.7 nm ($m=4$), in analogy with what previously reported in similar systems.\cite{tsymbalprl}

\begin{figure}
\begin{center}
\includegraphics[width=6.5cm]{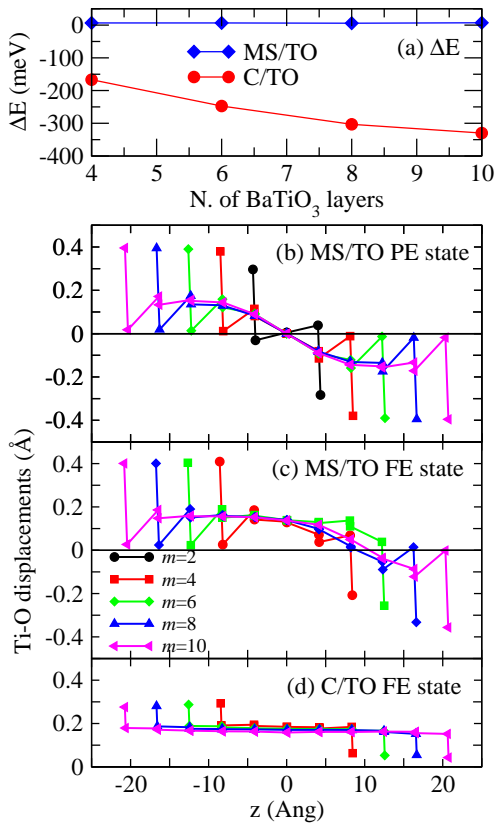}
    \caption{a) Difference of total energy between FE and PE atomic configurations for  MS/TO and C/TO interface structure vs number of BTO layers $m$ .  The relative Ti-O displacements in BTO vs the $z$ coordinate in the unit cell are shown for different $m$ in 
    (b) MS/TO with PE state; 
    (c) MS/TO with FE state; (d) C/TO with FE state.}
      \label{fig2}
\end{center}
\end{figure}

Figure \ref{fig2} a) shows the energy difference between the PE and FE states.
In the MS-termination, 
the PE state is more stable than the FE state,  although
the energy difference is comparable to our numerical uncertainty ($<$ 7 meV). This suggests that in the 
MS/TO junction, we cannot exclude a  FE state to be stabilized; for this reason, we will keep this latter system into account in the discussion of the electronic structure (see below).
As for the C/TO-termination, 
the FE state is strongly stable ($>$ 150 meV).  
The different behaviour of C/TO and MS/TO can be fully ascribed to the interface geometry:
{\em i)} In the MnSi-TiO$_{2}$ interface, because of the larger covalent radius of Mn with respect to the Si ion,
the bond length between O and Si atom is comparably shorter (1.78 \AA{} for $m=4$, PE state)
than that between O and Mn atom (1.96 \AA{} for same structure).
Therefore,  the O buckling strongly depends on the site:
the O atom on top of Si is  (relatively) displaced {\em towards} the interface whereas 
the O atom on top of Mn is  (relatively) displaced {\em away} from the interface.
This asymmetric bonding pattern prevents the spontaneous Ti-O FE polarization when 
the interface bonding energy  overcomes bulk ferroelectricity in BTO.
{\em ii}) In the C/TO interface, every O atom is atop of interstitial site of Co layer and 
the {\em asymmetrical} effect - just pointed out in {\em i}) - from subsurface MnSi layer is negligible
(the distance between O and Si atom is 3.94 \AA{} and the distance between O and Mn atom is 3.97 \AA{}).
As a result, the FE polarization becomes dominant
and every O atom is displaced in the same direction. 

This is quantified in Fig. \ref{fig2} where we show the relative displacements between Ti and O atoms in BTO. 
In the MS/TO interface, the relative Ti-O displacement induced by the interface bonding 
is about 0.4 \AA{} (irrespectively of $m$),   three times larger than  the FE Ti-O displacement in bulk BTO ( $\sim$ 0.125 \AA ).
In this case, O atoms at both sides of the interface are {\em pinned} and, inevitably, at one of the interfaces this  counteracts the Ti-O FE displacement pattern (see fig.\ref{fig1}). 
As a result, an interface domain wall  \cite{duan} appears in the FE state (cfr. Fig.\ref{fig2} c):
there is an overall net polarization in the insulating side, but the latter is divided into two regions with different thicknesses with oppositely oriented polarization.
 The situation is remarkably different at
the C/TO interface: the Ti-O displacement induced by interface bonding is much smaller ($<$ 0.2\AA).
The barrier between PE and FE state is then soon overcome and the FE state is stabilized as  the BTO thickness increases (at $m\ge4$).

\begin{figure}[htbp]
\includegraphics[width=7.cm]{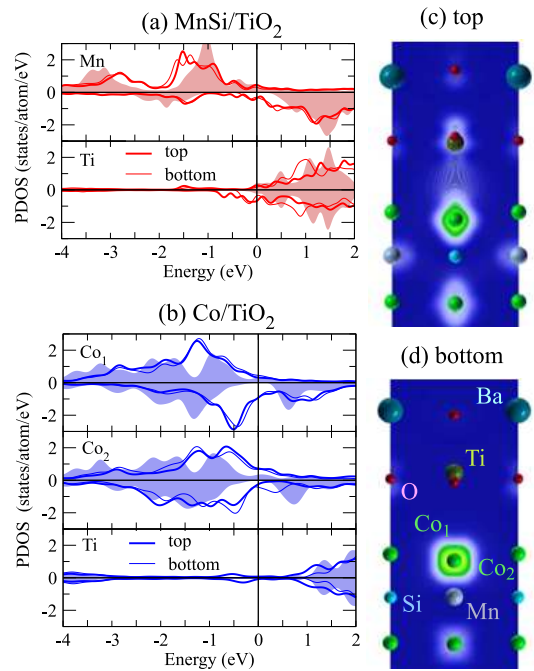}
    \caption{DOS for   
    (a) MS/TO interface with FE state (top and bottom panel show Mn and Ti $d$ states, respectively) and (b) C/TO interface  (top, middle and bottom panel show Co$_1$, Co$_2$ and Ti $d$ states). 
    The shaded area shows the atom at the middle ``bulk" layer in between interfaces.
    Thick (thin) solid lines show the DOS for the atom closest to 
    top (bottom) interface. c) and d) Minority spin charge density  (in arbitrary units) plotted in the plane which includes Co-Ti bond at the C/TO top and bottom interface, respectively
(Density calculated in the [-0.5 ; 0.0]-eV energy range with respect to $E_F$).}
      \label{fig4}
\end{figure}

In order to discuss half-metalicity (HM) at the interface, 
we focus on the electronic and magnetic properties.
Fig. \ref{fig4} a) and b) show the density of states (DOS) projected on $d-$orbitals of 
atoms at both edges of the interface.
The definition of top/bottom interface follows Fig. 1: 
in order for ferroelectricity to develop, most of Ti (O) atoms are displaced toward the top (bottom) junction, with a resulting net polarization in BTO pointing to the top interface.
{\em i}) In MnSi-TiO$_{2}$ interface (cfr Fig.\ref{fig4} a),  unoccupied Mn $d-$state and Ti $d-$state are well hybridized so that considerable Ti minority spin states appear close to Fermi energy, $E_F$.
This effect is enhanced at the top interface due to the relatively short bond length between Mn and Ti atom ($d_{\mathrm{Mn-Ti}}^{\mathrm{top}}=2.73\AA$, $d_{\mathrm{Mn-Ti}}^{\mathrm{bottom}}=2.82\AA$) and results in a negative magnetic moment of Ti (cfr Table\ref{table1}).
\begin{table}
\caption{Magnetic moments ($\mu_{B}$) of atoms at top and bottom interfaces for a) MS/TO and b) C/TO terminations.
  From now on, it is defined that Co$_{1}$ (Co$_{2}$) atom is at the same $(x, y)$ coordinate as Ba (Ti) atom.}
\label{table1}
\begin{center}
\begin{tabular}{ccc}
 \begin{tabular}{cccc}
  (a)&middle&top&bottom \\
  Co$_{1}$&1.1&1.0&1.0 \\
  Co$_{2}$&1.1&1.1&1.1 \\
  Mn        &2.9&2.9&3.1 \\
  Ti          &0.0&-0.3&-0.2
 \end{tabular}
 &
 &
 \begin{tabular}{cccc}
  (b)&middle&top&bottom \\
  Co$_{1}$&1.0&1.3&1.1 \\
  Co$_{2}$&1.0&-0.1&0.5 \\
  Mn        &2.8&2.4&2.5 \\
  Ti          &0.0&-0.05&0.0
 \end{tabular}
 \end{tabular}
\end{center}
\end{table}%
This situation is similar to Fe/BaTiO$_{3}$ study.\cite{tsymbalprl}
Although HM is no more preserved in the interface,  minority spin states which occur at the interface
 are efficiently screened in the Heusler region,
where the subsurface Co state (not shown) keeps its bulk HM.  It is instructive to compare our results with a recent ab-initio work focused on the pure CMS surface in the standard MS termination \cite{scheffler}. In both cases there is a loss of HM as well as an enhancement of the Mn magnetic moment with respect to the bulk, as expected from the rehybridization of Mn due to reduced number of Co. However, in the surface case, this results in a surface band crossing $E_F$ with mainly Co $d$ character; on the other hand, in our case the loss of HM is largely due to the Mn-Ti hybridization, therefore being mainly an ``interface-induced" effect. {\em ii}) At the Co$_{2}$-TiO$_{2}$ interface, 
the top- and bottom- surface Co$_{2}$ states widely differ and 
they are also different from the the bulk state (see in particular the different DOS in proximity to $E_F$)
The corresponding difference in magnetic moment (-0.1 $\mu_{B}$ at top, 0.5 $\mu_{B}$ at bottom) is remarkable.
This suggests that, if an electric field is applied to this junction so as to switch the FE polarization, 
the magnetic moment of Co atom would flip accordingly, with possible technological implications.
We note that, due to the different location in energy, the hybridization between Co and Ti $d-$states is rather small to cause this large ME effect.
Therefore, its origin may be better explained by a ``structural'' effect, due to different surface Co buckling. 
To gain further insight on this same issue, we show in
Fig. \ref{fig4} c) and d) the minority-spin charge density plots of the Co$_{2}$-TiO$_{2}$ interface in the energy region of the ``bulk"  CMS gap.
When comparing both interfaces, we observe that 
the ferroelectrically displaced Ti atom approaches Co atom  at the top- much more than at the bottom-interface 
 ($d_{\mathrm{Co-Ti}}^{\mathrm{top}}=2.43$ \AA, $d_{\mathrm{Co-Ti}}^{\mathrm{bottom}}=2.76$ \AA).
Although Co and Ti don't show a large hybridization, these large structural differences result in  a dramatic change on the electronic hybridization in the Heusler side  and, in turn, on the charge surrounding the Co atom:  the Co $3d$ electron orbital 
$t_{2g}$  like shape at the bottom interface is turned into an $e_g$-like shape at the top interface.  

In summary, our first-principles calculations point to  remarkable magnetoelectric effects at the \cms/BaTiO$_3$ interface between mainstream ferromagnetic half-metal and  ferroelectric. 
The atomic termination is found to profoundly affect the junction properties: in the MS/TO case the FE state in the BTO side energetically competes with a PE ground-state, whereas for the C/TO termination ferroelectricity is clearly stabilized. Our results evidently show that, for technological applications such as multiferroic tunnel junctions,\cite{kohl} not only should  the constituent materials be cautiously chosen, but also their atomic termination should be carefully engineered for optimized performances. In both MS/TO and C/TO terminations our observed ME effect shows a different origin with respect to  previous reports on, for example, BTO/CoFe$2$O$_4$\cite{schlom}, where magnetoelectricity is mediated by ``strain" induced by a piezoelectric material in contact with a magnetostrictive compound. In particular,  in the MS/TO  case the hybridization between unoccupied  transition-metal and Ti $d$ states is mainly responsible for magnetoelectricity, similar to what previously reported for Fe/BaTiO$_3$. However at the C/TO junction a different mechanism is proposed to explain the ME: the different atomic geometries at the two inequivalent interfaces (in turn related to FE displacements) profoundly affects the magnetic moment and order of the energy levels at the Co site, leading to a "structure-mediated" ME effect.

\end{document}